\documentclass{article}
\usepackage[left=3cm,right=3cm,top=3cm,bottom=3cm]{geometry} 
\usepackage{amsmath} 
\usepackage{authblk}
\usepackage{amsmath, amssymb, enumerate, amsthm}

\usepackage{graphicx,psfrag,epsf}
\usepackage{graphicx,hyperref}
\usepackage{amssymb}
\usepackage{amsmath}
\usepackage{multirow}
\usepackage{float}
\usepackage{mathtools}
\usepackage{xcolor}
\mathtoolsset{showonlyrefs=true}
\usepackage{natbib}
\usepackage{subfig}
\usepackage{lipsum}
\usepackage{graphicx}
\usepackage{epsfig}
\usepackage{epstopdf}
\usepackage{xr}
\usepackage{fancyhdr,lipsum}
\makeatletter

\newsavebox{\@linebox}
\savebox{\@linebox}[3em][t]{\parbox[t]{3em}{%
		\@tempcnta\@ne\relax
		\loop{\underline{\scriptsize\the\@tempcnta}}\\
		\advance\@tempcnta by \@ne\ifnum\@tempcnta<48\repeat}}

\usepackage{amssymb}

\begin{document}
	
	\title{CircSpaceTime: an R package for spatial and spatio-temporal modeling of
Circular data
	}  
	\author{{Giovanna Jona Lasinio$^{a}$, Mario Santoro$^{b}$, Gianluca Mastrantonio$^{c}$}\\
{$^{a}$Sapienza University of Rome, P.le Aldo Moro 5, 00185 Rome, ITALY; $^{b}$IAC CNR, Via dei Taurini, 19, 00185 Rome, ITALY; $^{c}$Polytechnic of Turin, ITALY, Corso Duca degli Abruzzi, 24, 10129 Turin,ITALY}
}

%

	\date{}

	\maketitle

			\begin{abstract}

			CircSpaceTime is the only R package currently available that
implements Bayesian models for spatial and spatio-temporal interpolation
of circular data. Such data are often found in applications where, among
the many, wind directions, animal movement directions, and wave
directions are involved. To analyze such data we need models for
observations at locations \(\mathbf{s}\) and times \(t\), as the
so-called geostatistical models, providing structured dependence assumed
to decay in distance and time. The approach we take begins with Gaussian
processes defined for linear variables over space and time. Then, we use
either wrapping or projection to obtain processes for circular data. The
models are cast as hierarchical, with fitting and inference within a
Bayesian framework. Altogether, this package implements work developed
by a series of papers; the most relevant being
\citet{Jona2013, wang2014, mastrantonio2015b}. All procedures are
written using Rcpp. Estimates are obtained by MCMC allowing
parallelized multiple chains run. The implementation of the proposed
models is considerably improved on the simple routines adopted in the
research papers. As original running examples, for the spatial and
spatio-temporal settings, we use wind directions datasets over central
Italy.
		\end{abstract}

		\maketitle

\hypertarget{intro}{%
\section{Introduction}\label{intro}}

In the last ten years the interest in circular data has received renewed
attention, with new theoretical results and models
\citep[for a recent and extended review of both theory and applications see][]{ley2017, ley2019}.
Most of the existing literature, before 2012, dealt with the development
of univariate distributions on the circle, regression type-models
including both circular and linear variable
\citep[fundamental textbooks are][]{mardia72, Merdia1999, Jammalamadaka2001}.
\citet{coles98} addressed circular time series and briefly sketched
possible extension to more complex dependence structures, such as spatial
and spatio-temporal dependence. However, an under-investigated area in
directional data was the development of space and space time processes,
where directional observations are taken at spatial locations, usually
over time as well, and proximity in space and time affects dependence
between directions. Such data are typical of the world of geostatistical
modeling \citep{Banerjee2014}, were structured dependence in space and
time is introduced. Here, we review work developed by the authors
together with other statisticians in a series of papers published over
the last few years focusing on spatial and spatio-temporal modeling and  we introduce a new formalisation of these models. We
 illustrate two case studies involving wind directions that
have not been previously analyzed as spatial and/or spatio-temporal
processes. In a purely spatial framework, we model wind direction in
order to interpolate them in space on a finer grid than the one
observed. While the aim of the space-time modeling exercise is the
(forward) prediction of an unobserved time point. All modeling efforts
are cast as hierarchical models, with fitting and inference within a
Bayesian inference framework. The hierarchical specification is vital;
latent variables are introduced to facilitate passage from linear
variables to circular. The Bayesian framework is arguably most
attractive for this setting. We obtain full posterior inference
including uncertainties, we avoid potentially inappropriate asymptotics,
and we enable routine prediction (kriging). Furthermore, model fitting
is straightforwardly implemented using Markov chain Monte Carlo
\citep{Banerjee2014}. Specifically, Gibbs and Metropolis sampling
provide loops, updating parameters given latent variables and data and,
then, updating latent variables given parameters and data
\citep{Jona2013}. The main novelty of this work is the implementation of
the models we are going to discuss, which has been highly improved on
the original settings associated to the research papers
\cite{Jona2013,Wang2013,wang2014, mastrantonio2015b}. In the package the
Matern, Gaussian and exponential covariance functions are available for spatial models
(only the exponential was explored in the research papers). All
estimations are obtained through MCMC and here we implemented the
algorithm shown in \cite{andrieu2008}, to speed up
estimation of structured covariance functions. Further we are adding the
possibility of running multiple chains in parallel using the
doParallel package.

None of the existing packages for circular and directional data  deal with spatial or spatio-temporal interpolation, that is
the main objective of CircSpaceTime see
Section \ref{pacchetti} for a listing of such packages and a short discussion of their
features). The package is currently
available from CRAN
(\url{https://CRAN.R-project.org/package=CircSpaceTime}).

There are different approaches to specify valid circular distributions,
see for example \cite{Jammalamadaka2001}, here we focus on the two
methods that allow to build a circular distribution starting from a
linear one, namely the wrapping, and the projection. Both revealed to be
useful in the definition of spatial and spatio-temporal data modeling,
see for example \cite{mastrantonio2015c} and \cite{wang2014}. Under both
methods, the resulting distribution has a complex functional form but
introducing suitable latent variables, the joint distribution of
observed and latent variables, in a fully Bayesian framework, is really
easy to handle.

\hypertarget{pacchetti}{%
\subsection{Existing software for circular and directional data}\label{pacchetti}}

There exist several software libraries  dealing with circular data, most of them are developed under the R project; in this framework the best
known are circular \citep{circular} and CircStats
\citep{circstats}, both implementing inference for univariate data as
described in \cite{Jammalamadaka2001}. Another recent set of functions
specifically devoted to wrapped distributions is Wrapped
\citep{Nadarajah:2017aa}. The package computes the probability density
function, cumulative distribution function, quantile function and many
more features for several (about fifty) univariate wrapped
distributions. A very interesting set of functions is implemented in
CircSizer \citep{Oliveira2014} where a non-parametric approach is
adopted. Based on scale-space ideas, CircSiZer presents a graphical
device to assess which observed features are statistically significant,
both for density and regression analysis.\\
In the likelihood and univariate framework, we find a good book
on circular data in R \citep{arthurbook} where many nice examples
can be found and the narrative of the topic makes easy to learn how to
run inferences on univariate data. In 2013 the first version of the
package isocirc was presented \citep{Barragan:2013}, making
available functions to perform constrained inference using isotonic
regression \citep{rueda2009, FERNANDEZ2011}. The CircOutlier
\citep{CircOut} collects functions to detect outliers in
circular-circular regression as proposed in \citet{Abuzaid2013}. More
recent is the Directional package \citep{directional}, mostly linked
to the textbook by \citet{Merdia1999}. A series of wrapper functions to
implement the 10 maximum likelihood models of animal orientation,
described by \citet{SCHNUTE1992}, are included in the CircMLE. The
proposals in \citet{zimmermann2017} are presented in circumplex.

Bayesian estimation for univariate regression models is implemented in
bpnreg \citep{bpnreg} that presents models developed in
\cite{nunez2014} and \cite{cremer2017}. Again in the Bayesian framework
the work in \citet{MULDER2017} is implemented in circglmbayes.

Dependent and multivariate circular data are often found in applications
\citep[see][for recent developments]{ley2019}. To handle them in a
likelihood framework we can refer to the package CircNNTSR
\citep{Fernandez2016}, that implements functions to plot, fit by maximum
likelihood, and simulate models based on non-negative trigonometric sums
for circular, multivariate circular, and spherical data.

Outside the R project we find commercial softwares too, such as Oriana\footnote{https://www.kovcomp.co.uk/index.html} for Windows performing basic statistics, testing and graphics or NCSS, LLC Statistical Software that has a section devoted to circular data, always performing simple basic statistics. In the Matlab environment we find the Circular Statistics Toolbox (Directional Statistics) \citep{berens2009} again performing descriptive statistics, basic testing (one sample and multiple samples) and graphics.

There are no packages or software performing spatial interpolation nor space-time
prediction for directional or circular data.
 
\hypertarget{lineTocirc}{%
\section{From Linear to Circular models}\label{lineTocirc}}
We introduce, in both univariate and
multivariate settings, the two approaches that we exploit to built
distributions/processes for circular variables, from probability
distributions with support the real line; we start with
\emph{linear random variables} to obtain
\emph{circular random variables}. The two methods are totally general
and can be applied to linear variables with any distribution. We focus
on the Gaussian case, since the resulting models are highly flexible and
they are those implemented in CircSpaceTime.

\hypertarget{wrap}{%
\subsection{The wrapping approach}\label{wrap}}

With the wrapping approach, the idea is to ``wrap'' the density of a
linear variable around the unit circle, obtaining than a circular
density. In more details, let \({Y} \in \mathbb{R}\) be a linear random
variable with pdf \(f_{Y}(\cdot|\boldsymbol{\psi})\), where
\(\boldsymbol{\psi}\) is a vector of parameters. We obtain a circular
random variable using the following transformation:
\begin{equation} \label{eq:wrap} \Theta = {Y} \text{ mod } 2 \pi \in [0,2 \pi).
\end{equation} The pdf of \(\Theta\) is then
\begin{equation} \label{funzioni} f_{\Theta}(\theta|\boldsymbol{\psi}) =\sum_{k= -\infty}^{\infty}f_Y(\theta+2\pi k|\boldsymbol{\psi}).
\end{equation}

It is easy to find the relation between \(Y\) and \(\Theta\) that is the
following: \(Y = \Theta+2 \pi K\), where \(K\) is the so-called
\emph{winding number}. Equation \eqref{funzioni} wraps the density
\(f_Y(\cdot|\boldsymbol{\psi})\) around the unit circle and then
\(\Theta\) is called the \emph{wrapped} version of \(\mathbf{Y}\) of
period \(2 \pi\). If \({Y}\) is normally distributed, then \(\Theta\)
follows a \emph{wrapped normal} (WN) distribution.

The evaluation of \eqref{funzioni} is not easy since it involves an
infinite sum, possibly, making inference based on the wrapped
distribution a really complex task. Following \citet{Coles1998}, this
problem can be easily overcome if we consider \(K\) as a (latent) random
variable. We have then
\(f_{\Theta,K}(\theta,k|\boldsymbol{\psi})= f_Y(\theta+2 \pi k|\boldsymbol{\psi})\),
i.e.~the joint density of \((\Theta,K)\) is
\(f_Y(\theta+2 \pi k|\boldsymbol{\psi})\), that is the density \(Y\),
seen as function of \(\theta\) and \(k\). Notice that \eqref{funzioni}
can be seen as a marginalization over \(K\) of the joint density of
\((\Theta,K)\). The conditional distribution of \(K\), that is needed
for the implementation of the MCMC, is easy to handle since it is
proportional to \(f_Y(\theta+2 \pi k|\boldsymbol{\psi})\). This result
shows that it is much easier to work with the joint density of
\((\Theta,K)\), with respect to the marginal of \(\Theta\).

The wrapping approach can be easily extended to a multivariate setting,
see for example \citet{Jona2013}. In details, let
\(\mathbf{Y} =(Y_1,\dots, Y_n)^{\prime}\) be a \(n\)-variate vector with
pdf \(f_{\mathbf{Y}}(\cdot| \boldsymbol{\psi})\). We have then that
\(\boldsymbol{\Theta}=(\Theta_1,\dots, \Theta_n)^{\prime}\), with
\begin{equation} \label{eq:theta_t} 
\Theta_i = {Y}_i \text{ mod } 2 \pi , i=1,\dots, n,
\end{equation} is a vector of circular variables with density
\begin{equation} \label{eq:wmulti}
f_{\boldsymbol{\Theta}}(\boldsymbol{\theta}|\boldsymbol{\psi}) =\sum_{k_1=
-\infty}^{\infty}\dots \sum_{k_n= -\infty}^{\infty}
f_{\mathbf{Y}}(\boldsymbol{\theta}+2 \pi \mathbf{k}|\boldsymbol{\psi}).
\end{equation}

Extending the reasoning applied to the univariate setting, we can easily
find that the full conditional of \(\mathbf{K}\) is proportional to
\(f_{\mathbf{Y}}(\boldsymbol{\theta}+2 \pi \mathbf{k}|\boldsymbol{\psi})\)
and the joint density of \((\boldsymbol{\Theta},\mathbf{K})\) is
\(f_{\mathbf{Y}}(\boldsymbol{\theta}+2 \pi \mathbf{k}|\boldsymbol{\psi})\).
If we assume that \(\mathbf{Y}\) is normally distributed, this means
that the joint density of \((\boldsymbol{\Theta},\mathbf{K})\) is

\begin{equation} \label{eq:sim1}
\phi_{n}(\boldsymbol{\theta}+2 \pi \mathbf{k}|\boldsymbol{\mu}, \boldsymbol{\Lambda}),
\end{equation} where
\(\phi_n(\cdot|\boldsymbol{\mu},\boldsymbol{\Lambda})\) is the
\(n-\)variate normal pdf with mean vector and covariance matrix given by
\((\boldsymbol{\mu},\boldsymbol{\Lambda})\). Here again it is easy to
see that any type of model implementation is facilitated if
\(\mathbf{K}\) is introduced as a latent variable.

\hypertarget{projection}{%
\subsection{The projection approach}\label{projection}}

With the wrapping approach, from one linear variable we obtain a
circular one. The projection approach is quite different in this regard
since it requires two linear variables to obtain one circular. The basic
idea is to take the two variables, that can be seen as coordinates in a
Euclidean space, and express them as polar coordinates. The associated
angle is itself a random variables and it is circular.

Let \(\mathbf{Y}= (Y_1,Y_2)\) be a bivariate vector of linear variables
with pdf \(f_{\mathbf{Y}}(\cdot|\boldsymbol{\psi})\). The relation
between the angle \(\Theta\) and \(\mathbf{Y}\) is expressed by the
following: \begin{equation}
\tan(\Theta) =\frac{Y_2}{Y_1}. \label{eq:oo}
\end{equation} The circular variable can be then obtained using the
inverse tangent function but some care must be taken. The inverse
tangent has domain of length \(\pi\) while we want our random variable
to have domain \([0,2\pi)\). This is generally achieved by using the
\(\mbox{atan}^*\) inverse
\citep[see][pag.13 for a detailed definition]{Jammalamadaka2001}, that
takes into account the sign of the components of \(\mathbf{Y}\), to
identify to which of the 4 quadrants the circular variable belongs.
Between \(\Theta\) and \(\mathbf{Y}\) the following relation exists
\({\bf Y} = \begin{bmatrix} Y_{1}\\ Y_{2}\end{bmatrix} =R \begin{bmatrix}\cos\theta\\ \sin\theta\end{bmatrix} = R{\bf U},\)
with \(R= ||\mathbf{Y}||\), where the joint vector \((\Theta,R)'\) is
the representation in polar coordinates of \(\mathbf{Y}\). From equation
\eqref{eq:oo} is easy to see that distributions based on the projection
are not identifiable, since \(c\mathbf{Y}\) and \(\mathbf{Y}\), with
\(c>0\), gives rise to the same circular variables; an identification
constraint is then needed \citep{Wang2013}.

The pdf of \(\Theta\) is \begin{equation} \label{eq:dproj} 
f_{\Theta}(\theta|\boldsymbol{\psi}) =
\int_{\mathbb{R^+}} r f_{\mathbf{Y}} ((r\cos(\theta),r
\sin(\theta))^{\prime}|\boldsymbol{\psi}) d r,
\end{equation} that is obtained by finding first the joint density of
\((\Theta,R)\), which is \begin{equation}
r f_{\mathbf{Y}} ((r\cos(\theta),r
\sin(\theta))^{\prime}|\boldsymbol{\psi}),
\end{equation} and then, a marginalization over \(R\), gives the density
of \(\Theta\). The integral in equation \eqref{eq:dproj} is not easy to
solve and, even when a closed form exists, the resulting pdf has a
complicated functional structure. For example, if we assume
\(\mathbf{Y} \sim N_2(\boldsymbol{\alpha},\boldsymbol{\Xi})\), with
\(\boldsymbol{\alpha}=(\alpha_1,\alpha_2)^{\prime}\) and
\begin{equation}
\boldsymbol{\Xi} =\left( \begin{array}{cc} \sigma^2& \tau \sigma \\
\tau \sigma  & 1 \end{array} \right), \label{eq:xi}
\end{equation} where \((\boldsymbol{\Xi} )_{2,2}=1\) is needed for
identifiability purpose, the density of \(\Theta\), computed using
\eqref{eq:dproj}, is \begin{equation}
f_{\Theta}(\theta|\boldsymbol{\psi}) =
\frac{\phi_2(\boldsymbol{\alpha}|\mathbf{0}_2,\boldsymbol{\Xi})+a D(\theta)
    \Phi(D(\theta)|\mathbf{0}, \mathbf{I}_2) \phi(a C(\theta)^{-\frac{1}{2}}
    (\alpha_1 \sin(\theta))-\alpha_2 \cos(\theta)   )  }{C(\theta)}, \label{eq:2}
\end{equation} where \(\Phi_n(\cdot|\cdot,\cdot)\) is the \(n-\)variate
cumulative Gaussian density function, with \begin{align} 
&a=  \left(\sigma \sqrt{1-\tau^2}\right)^{-1},\\
&C(\theta) = a^2\left( \cos^2(\theta) +\sigma^2\sin^2(\theta) -\tau
\sigma \sin (2 \theta)  \right),\\ 
&D(\theta) = a^2
C(\theta)^{-\frac{1}{2}} \left(  \alpha_1    ( \cos (\theta)-
\tau \sigma \sin (\theta))+  \alpha_2 \sigma   (\sigma \sin (\theta)- \rho
 \cos (\theta))    \right). 
\end{align}

Under the normal assumption, we have that the joint density of
\((\Theta,R)\) is \begin{equation}
r \phi_2((r\cos(\theta),r
\sin(\theta)|\boldsymbol{\alpha},\boldsymbol{\Xi}) .\label{eq:ssa}
\end{equation} Equation \eqref{eq:ssa} is less complex than \eqref{eq:2}
but, as for the wrapping approach, to work with it we need to introduce
a latent variable, that is \(R\).

The extension of the projection approach to multivariate variable is
straightforward. We take \(\mathbf{Y}\) as a \(2n-\)variate linear
variable, and we build the \(n-\)variate vector of (projected) circular
variables through the following transformation:
\begin{equation} \label{eq:theta_t2}  
\Theta_i =\mbox{atan}^*\left(\frac{Y_{2i }}{Y_{2i- 1}}\right), i=1,\dots,n,
\end{equation} i.e., the \(2n\) elements of \(\mathbf{Y}\) are grouped
in \(n\) sets, each of them containing only two values, and where one
element of \(\mathbf{Y}\) can only belong to one of the \(n\) sets. Each
set gives rise to a circular variable. The density of the vector
\(\boldsymbol{\Theta} = (\Theta_1,\dots ,\Theta_p)^{\prime}\) is then
\begin{equation}
f_{\boldsymbol{\Theta}}(\boldsymbol{\theta}|\boldsymbol{\psi}) =
\int_{\mathbb{R^+}} \dots \int_{\mathbb{R^+}}  \prod_{i=1}^n r_i f_{\mathbf{Y}}
(\mathbf{y}(\boldsymbol{\theta},\mathbf{r})|\boldsymbol{\psi})  d r_1 \dots d r_n, \label{eq:s}
\end{equation} where \(r_i= || (y_{2i-1},y_{2i})^{\prime} ||\) and
\(\mathbf{y}(\boldsymbol{\theta},\mathbf{r})\) is used to indicated that
\(\mathbf{y}\) must be seen as function of \(\boldsymbol{\theta}\) and
\(\mathbf{r}=(r_1,\dots,r_n)\). Even assuming a normal density for
\(\mathbf{Y}\), it is not possible to write in closed form \eqref{eq:s}.
As for the univariate case, it is much easier to work with the joint
density of \((\boldsymbol{\Theta},\mathbf{R})\) since, if for example a
normal distribution of \(\mathbf{Y}\) is assumed, it is then
\begin{equation}\label{eq:sim2}
  \prod_{i=1}^n r_i \phi_{2n}
(\mathbf{y}(\boldsymbol{\theta},\mathbf{r})|\boldsymbol{\mu}, \boldsymbol{\Lambda}),
\end{equation} where, as in equation \eqref{eq:sim1},
\((\boldsymbol{\mu}, \boldsymbol{\Lambda})\) are the mean and covariance
matrix.

\hypertarget{differences-between-the-two-approaches}{%
\subsubsection{Differences between the two
approaches}\label{differences-between-the-two-approaches}}

There are considerable differences between the two approaches discussed
above. The wrapping creates circular distributions that are in general
similar to their counter-parts on the real line. For example, the
wrapped normal is still unimodal and symmetric. Moreover, the circular
mean and variance
\citep[see][for the definitions and details]{Jammalamadaka2001}, are
simple functions of the mean and variance of the linear variable.

On the other hand, even under the normal assumption, the projection
approach creates a distribution that often shows very different
properties from its linear partners. Again in the Gaussian setting, the
projected normal can be bimodal, asymmetric and antipodal. Moreover,
only in few special cases, exists a closed form expression of the mean
and variance of the projected distribution.

The main reason to propose these two approaches is that, in both, it is
easy to introduce dependence (spatial, temporal or both). The wrapping
gives results that are really easy to interpret in terms of phenomena
behaviour, while the projection is very useful when interpretation is
not central, and a highly flexible model is required
\citep{mastrantonio2015b}.

\hypertarget{spa}{%
\subsection{Spatio-temporal processes for circular
variables}\label{spa}}

A stochastic process can be defined through its finite-dimensional
distribution, i.e.~the distribution of an \(n-\)dimensional realization,
that has a multivariate pdf \citep{gelfand2010}. Since in the previous
sections we already described multivariate distributions for circular
variables, it is then easy to define circular processes. More precisely,
starting from a distribution for linear variables, we can use the
wrapping or the projection approach to obtain a multivariate circular
distribution and then, as a consequence, from an \(n-\)dimensional
realization of a linear process we can obtain the associated
\(n-\)dimensional realization of a circular one.

In particular, in the Gaussian cas, let
\({Y}(\mathbf{s})= \{ Y_{i}(\mathbf{s}) \}_{i=1}^p \in \mathbb{R}^p\) be
a \(p-\)variate Gaussian process (GP) defined over a \(s-\)dimensional
domain, i.e., \(\mathbf{s}\in \mathcal{S}\subset \mathbb{R}^d\), and let
\(\mathbf{y}\equiv (\mathbf{y}_1,\dots , \mathbf{y}_n)'\in \mathbb{R}^p \times\mathbb{R}^n\)
be the set of \(n\) realizations. The vector \(\mathbf{y}\) is then
normally distributed. Given \(\mathbf{Y}(\mathbf{s})\) we can easily
built the wrapped and the projected GP using \eqref{eq:theta_t} and
\eqref{eq:theta_t2}, respectively. More precisely, the former is
obtained as \[ \Theta(\mathbf{b}) = \mathbf{Y}(\mathbf{s}) \text{ mod
} 2 \pi. \] while the latter is given by \begin{equation} 
\Theta(\mathbf{s}) =
\mbox{atan}^*\left(\frac{Y_{2}(\mathbf{s})}{Y_{1}(\mathbf{s})}\right).
\end{equation} The same transfromation, applied to \(\mathbf{y}\), give
the realization of the wrapped and projected GP.

For both, the spatial or space-time representations are obtained by
considering a spatially or spatio-temporally structured covariance
matrix.

\hypertarget{the-implemented-models}{%
\section{The implemented models}\label{the-implemented-models}}

\begin{table}[t]
  \centering
  \begin{tabular}{c|c|c}
    \hline \hline
    Name & Function & Parameters\\\hline
    Mat\`ern &$\frac {2^{1-\nu }}{\Gamma (\nu )}\left({\sqrt {2\nu }}{\frac {h_{\text{sp}}}{\rho }}\right)^{\nu }K_{\nu }\left({\sqrt {2\nu }}{\frac {h_{\text{sp}}}{\rho }}\right)$& $\nu, \rho$\\ Exponential& $\exp\left(-\rho h_{\text{sp}}  \right)$ & $\rho$\\
    Gaussian&$\exp\left(-\rho^2 h_{\text{sp}}  \right) $ & $\rho$\\
    Gneiting (spatio-temporal)& $\frac{1}{\rho_{t}h_{\text{t}}^{2}+1}\exp\left(-\frac{\rho_{sp}h_{\text{sp}}}{(\rho_{t}h_{\text{t}}^{2}+1)^{\frac{\eta }{2}}}\right)$& $\rho_{t}, \rho_{sp}, \eta$  \\
    \hline \hline
  \end{tabular}
  \caption{Correlation functions implemented. $h_{\text{sp}}$ and $h_{\text{t}}$ are, respectively, spatial and temporal distance.} \label{tab:covfun}
\end{table}

The basic linear model we adopt is the following: \begin{align}
\mathbf{Y}(\mathbf{s})& =
\boldsymbol{\alpha}+\boldsymbol{\omega}(\mathbf{s}),\label{eqref:y}\\
\boldsymbol{\omega}(\mathbf{s}) & \sim GP_{p}(\mathbf{0}_p, C(
\mathbf{h};\boldsymbol{\varphi}  )  \otimes \boldsymbol{\Xi}),
\end{align}

where \(\boldsymbol{\alpha} \in \mathbb{R}^p\) is a mean vector,
\(\boldsymbol{\omega}(\mathbf{s})\) is a zero mean \(p-\)variate GP and
\(\otimes\) is the usual Kronecker product. The correlation function of
the GP is \(C( \mathbf{h};\boldsymbol{\varphi} )\) and it depends on
parameter \(\boldsymbol{\varphi}\) and the vector of distances
\(\mathbf{h}\), that contains the spatial and temporal distances
\(h_{\text{sp}}\) and \(h_{\text{t}}\) if \(d=3\), and only
\(h_{\text{sp}}\) if \(d=2\). Under \(p=1\) the model \eqref{eqref:y} is
used to build a wrapped GP while \(p=2\) is used for the projected.
Table \ref{tab:covfun} shows the correlation functions and Table
\ref{tab:priors} the choice of prior distributions available in the
package CircSpaceTime. The parameter \(\boldsymbol{\Xi}\) is defined
as in equation \eqref{eq:xi} if \(p=2\), and coincide with \(\sigma^2\)
if \(p=1\).

\hypertarget{model-implementation}{%
\subsection{Model implementation}\label{model-implementation}}

\begin{table}[t]
\centering
\begin{tabular}{l|c |c}
\hline\hline
Parameter&Wrapped Normal& Projected Normal \\\hline\hline
Spatial and temporal decay ($\rho, \rho_{sp},\rho_t$) & uniform &uniform \\
$\sigma^2$&inverse gamma & inverse gamma\\
Separability parameter ($\eta=$ \texttt{sep\_par}) & Beta& Beta  \\
mean parameters ($\alpha$) & Wrapped Gaussian& Gaussian\\
correlation between components ($\tau$)&&uniform\\
\hline\hline
\end{tabular}
\caption{Available prior distributions }\label{tab:priors}
\end{table}

As we stated in Section \ref{lineTocirc}, the multivariate wrapped and
projected normal have densities too complex to be used for model
inference but, if we introduce as latent observation the vector
\(\mathbf{k}\), for the wrapped normal, and \(\mathbf{r}\) for the
projected, inference through MCMC algorithm is straightforward.

The key element here is that, once the latent variables are obtained,
the full conditionals of the model parameters are exactly the same as if
we were working with observed linear variables. This is due to the
functional form of the likelihood of circular and latent variables, see
\eqref{eq:sim1} and \eqref{eq:sim2}, which gives the same contribution
to the full conditionals of the parameters, as if the data has a
multivariate normal density.

Due to the priors available in the package, see Table \ref{tab:priors},
the update of \(\boldsymbol{\alpha}\), i.e.~the mean parameter, is
obtained with a Gibbs sample since its full conditional is multivariate
normal under the projected model and wrapped normal under the wrapped
model. The parameters of the covariance structure are updated all
together, within a Metropolis step. To speed up convergence and to have
the possibility to choose the final acceptance rate, for the
multivariate proposal distribution we implemented an algorithm shown in
\cite[][algorithm 4]{andrieu2008}. In details, let
\(\mathbf{x}_{ad}^{b} \in \mathbb{R}^{n_{ad}}\) be a
\(n_{ad}\)-dimensional vector, drawn at the \(b-\)th iteration of the
MCMC algorithm, from a multivariate normal with mean
\(\boldsymbol{\mu}_{ad}^{b}\) and covariance matrix
\(\lambda_{ad}^b\boldsymbol{\Sigma}_{ad}^{b}+{0.0001}\mathbf{I}_{n_{ad}}\),
where \(n_{ad}\) is the number of parameters of the model covariance
matrix. Each element of \(\mathbf{x}_{ad}^{b}\) is used to propose one
of the covariance parameters using suitable transformation:
\(\exp({x_{ad,j}})\) for the parameters belonging to \(\mathbb{R}^+\)
and \((a+b\exp({x_{ad,j'}}) )/(1+\exp({x_{ad,j'}} ))\) for the ones
defined over \([a,b]\). Parameters \(\boldsymbol{\mu}_{ad}^{b}\),
\(\lambda_{ad}^b\) and \(\boldsymbol{\Sigma}_{ad}^{b}\) are updated to
guarantee selected acceptance-ratio \(d_{ad}\), via the following:
\begin{align}
\lambda_{ad}^{b+1} & = \exp{\left(   \log \lambda_{ad}^{b}+\frac{1}{b^{\xi_{ad}}}(\alpha_{mh}-d_{ad})\right)}\\
\boldsymbol{\mu}_{ad}^{b+1} & = \boldsymbol{\mu}_{ad}^{b}+\frac{1}{b^{\xi_{ad}}}(\mathbf{x}_{ad}^{*}-\boldsymbol{\mu}_{ad}^{b})\\
\boldsymbol{\Sigma}_{ad}^{b+1} &=\boldsymbol{\Sigma}_{ad}^{b}+\frac{1}{b^{\xi_{ad}}}\left( (\mathbf{x}_{ad}^{*}-\boldsymbol{\mu}_{ad}^{b})  (\mathbf{x}_{ad}^{*}-\boldsymbol{\mu}_{ad}^{b})'-\boldsymbol{\Sigma}_{ad}^{b}  \right)
\end{align} where \(\alpha_{mh}\) is the Metropolis-Hastings ratio,
\(\xi_{ad} \in (0,1)\) is a parameter that allows to choose how much the
last iteration influences the next one; \(\mathbf{x}_{ad}^{*}\) is equal
to \(\mathbf{x}_{ad}^{b}\) if the proposed vector is accepted otherwise
the previous iteration value (\(\mathbf{x}_{ad}^{b-1}\)) is kept. Notice
that the adaptation decreases with the iterations while the larger the
parameters \(\xi_{ad}\), the weaker is the action of adaptation. The
diagonal element \({0.0001}\mathbf{I}_{n_{ad}}\) is needed to always be
able to invert the covariance matrix avoiding ill defined matrices. More
details on how and which parameters can be selected are given in the
next section.

The latent discrete variable \(\mathbf{k}\) is updated component-wise
with Metropolis steps. In each one of them, given the current value
\(k_i\), the proposal is selected from the set \((k_i-1,k_i,k_i+1)\),
where each element has probability 1/3 to be selected. The latent
variable \(\mathbf{r}\) is also updated component-wise with Metropolis
steps, using the algorithm proposed by \cite{ROSENTHAL20075467}. In this
case, at the \(b-\)th iteration we propose a new value of \(r_i\) from a
log-normal with mean given by \(\log{r_i^{b-1}}\) and standard deviation
\(sd_{ad,i}^b\). Every \(n_{batch}\) iterations, \(sd_{ad,i}^b\) is
updated using the following: \begin{equation}
sd_{ad,i}^{b+1} = \exp\left( \log sd_{ad,i}^{b} +\frac{1}{b^{\xi_{ad}}}(\bar{\alpha}_{mh}-d_{ad})\right),
\end{equation} otherwise \(sd_{ad,i}^{b+1} = sd_{ad,i}^{b}\), where
\(\bar{\alpha}_{mh}\) is the mean Metropolis-Hastings ratio over the
past \(n_{batch}\) iterations.

In fitting the wrapped Gaussian models we center data (and all related
quantities) around \(\pi\). Outputs for the mean and the interpolated
values are given on the original scale, while the winding numbers are
those obtained centering around \(\pi\).

To improve performances we implemented everything in C++
\citep{wickham2015} and using Rcpp package we simplify the
integration between C++and R codes \citep{Eddelbuettel2011}. In
particular we used the RcppArmadillo package
\citep{Eddelbuettel2014}, that implements the Armadillo matrix library,
for its simplicity and elegance \citep{Eddelbuettel2017}, although the
RcppEigen is a bit faster \citep{Eddelbuettel2015}. For a fast
multiple chain estimations we use doParallel package
\citep{doParallel}.

\hypertarget{two-case-study-based-on-wind-directions}{%
\section{Two case study based on wind
directions}\label{two-case-study-based-on-wind-directions}}

In this section we illustrate all the proposed models trough real data examples. The code associated to the examples is available as supplementary material.
%
%
%
%
%
%
%

\hypertarget{dataexamples}{%
\subsection{Wind data in Central Italy}\label{dataexamples}}

We consider examples based on wind data downloaded from the Global
Forecast System (GFS) of the USA's National Weather Service (NWS)
(\url{https://www.ncdc.noaa.gov/data-access/model-data/model-datasets/global-forcast-system-gfs}).
Wind data are taken from NOAA/NCEP Global Forecast System (GFS)
Atmospheric Model collection. Geospatial resolution is 0.5 degrees
(approximately 50 km), and the wind is calculated for Earth surface, at
10 m. The function \texttt{wind.dl} from rWind library
\citep{Fernandez-Lopez:2018} facilitate the download of the data.

Downloaded data are arranged into dataframes with the following
information:

\begin{itemize}
\item  the time point expressed as date-time (\texttt{time })
\item the coordinates as decimal longitude and latitude (\texttt{lat, lon} )
\item  the   eastward and northward wind velocity components at 10m a.s.l. (m/s) (\texttt{ugrd10m, vgrd10m})
\item the  direction of the wind in degrees ( \texttt{dir}).
\item  the  wind velocity in m/s (\texttt{speed}).
\end{itemize}

Directions are transformed to radians for modeling purposes and rotated
of 180 degrees to get values in accordance with the usual wind rose.

We select two days: the 29th of June 2012 and the 29th of October 2018.
The first date was a lightly breezy sunny day over the entire central
Italy. Light breeze in the area implies a large variability in terms of
wind directions. The second date saw one of the worst storms in the last
century hitting Europe. Central Italy was strongly affected by strong
winds and loss of lives. In this meteorological condition winds are
usually coming from one direction and the variability is very low.

\begin{figure}[t]
\centering
\includegraphics[scale = 0.35,trim= 20 110 20 80]{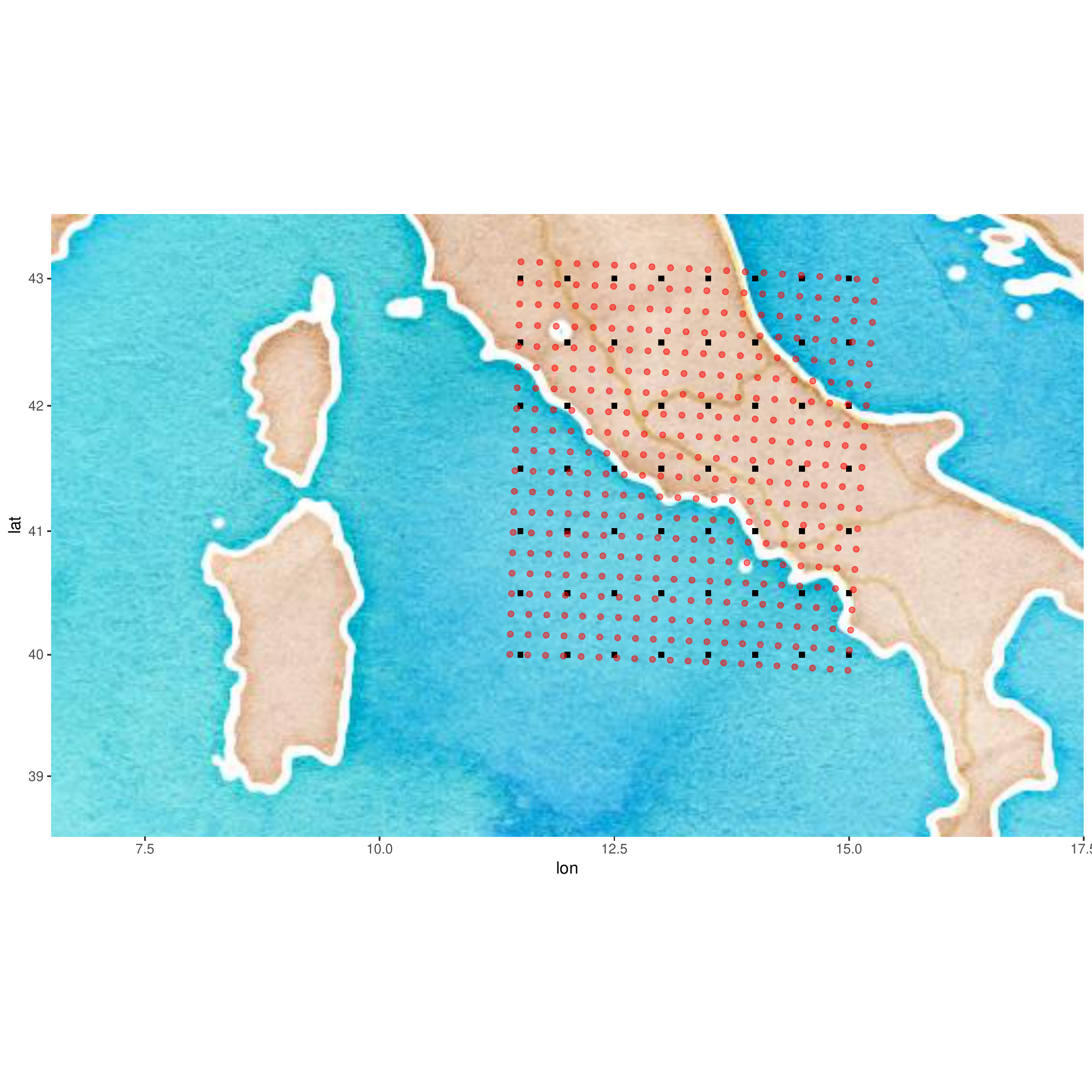}
\caption{Interpolation grid (red) and data points (black)}\label{fig:thegrid}
\end{figure}

We are interested in illustrating different modeling strategies dealing
with very different data variability. In the spatial setting, the aim of
the study is to provide directions on a grid with finer resolution (about 17 $\times$ 18 km) than the
about 50 km \(\times\) 50 km grid cells (see Figure \ref{fig:thegrid}) of the observations (56 observed points, 400 prediction points).
From June 29, 2012 we download data 3 hours apart, starting at
midnight and we choose the most variable time point for spatial interpolation. On October 29, 2018 we again download data  3 hours apart (from 3pm until 11pm),  we choose the  23:00 (11pm)  time point when
the peak of the storm hit the area, for spatial interpolation and we save the remaining for the spatio-temporal analysis.

\begin{figure}[t]
\centering
\includegraphics[scale = 0.35,trim= 20 20 20 20]{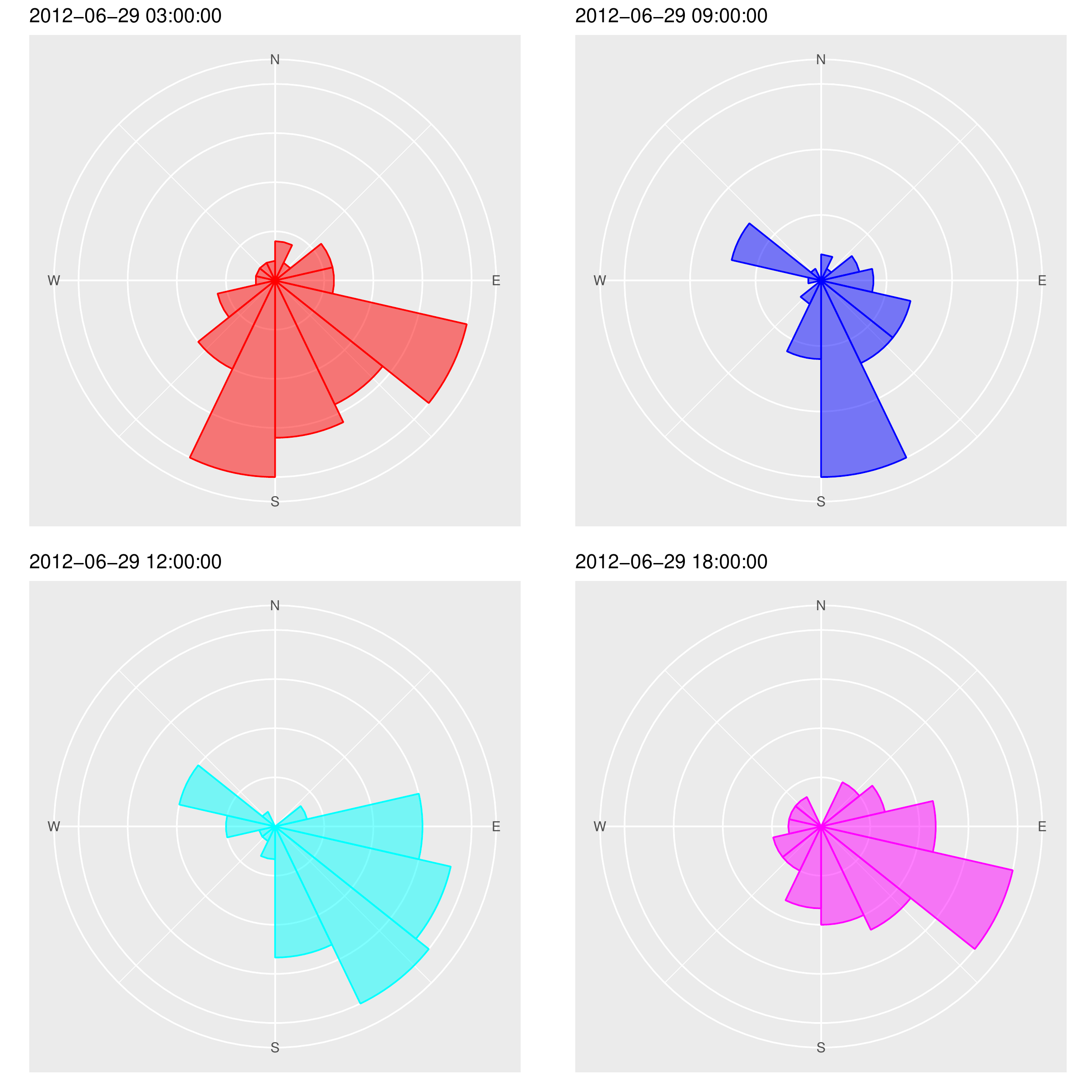}
\caption{Rose diagram of wind directions of June 29, 2012 at 4 time points }\label{fig:rose0}
\end{figure}

\begin{figure}[t]
\centering
\includegraphics[scale = 0.35,trim= 20 160 20 20]{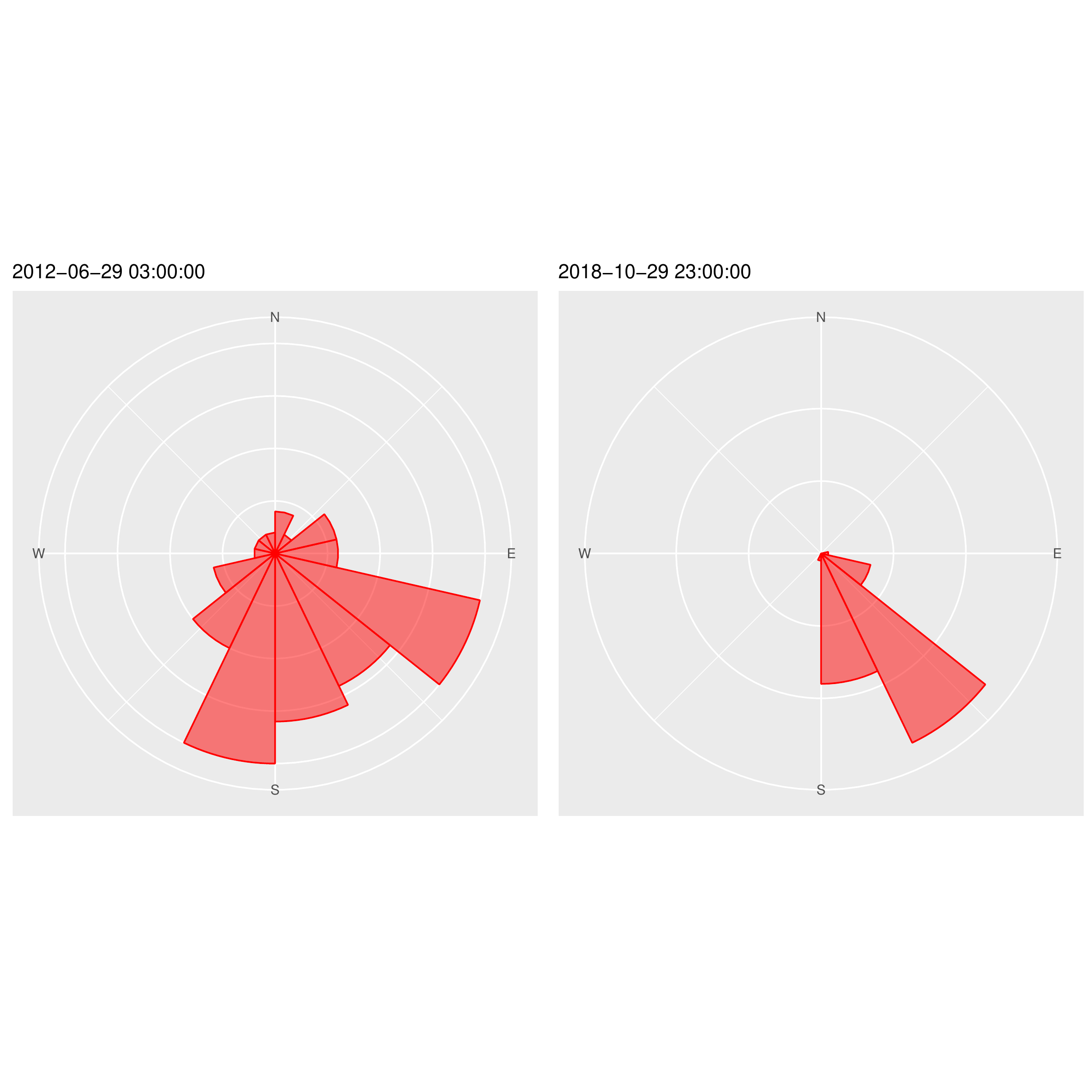}
\caption{Rose diagram of wind directions of June 29, 2012 at 3am and  October 29, 2018 at 23:00 (11pm) }\label{fig:bothdataset}
\end{figure}

\begin{figure}[t]
\centering
\includegraphics[scale = 0.55,trim= 20 240 20 220]{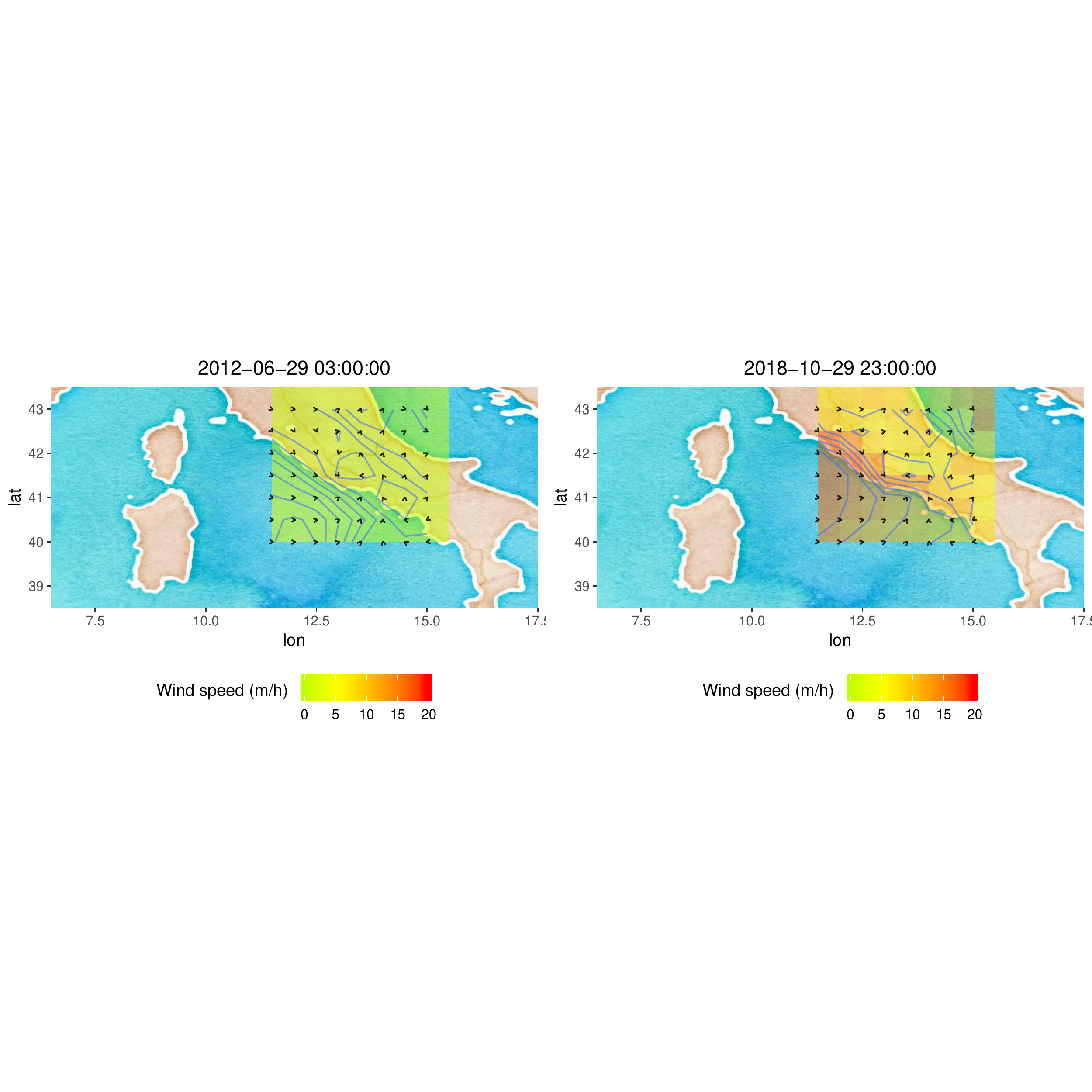}
\caption{Central Italy: wind directions on June 29, 2012 at 3am and October 29, 2018 at 11pm, the color scale follows the speed of the wind from green to red }\label{fig:map1}
\end{figure}

In Figure \ref{fig:rose0} we explore some of the time points on June 29,
2012. We choose the time slice at \(03:00\) as highly variable and
almost bimodal. We compare June 29, 2012 and October 29, 2018
variability in Figure \ref{fig:bothdataset}, the circular variance for
the first one is 0.48 and for the second dataset is 0.048, ten times
smaller.
In Figure \ref{fig:map1} we illustrate the observed wind directions together with the wind speed surface.

In what follows our strategy is to estimate and compare prediction from
both wrapped and projected normal models. Comparison is carried out by
computing the Average Prediction Error (APE) \cite[see][]{Jona2013} and
the circular continuous ranked probability score (CRPS)
\citep{grimit2006}. 

\hypertarget{the-29th-of-june-data-spatial-interpolation-of-highly-variable-data}{%
\subsection{The 29th of June data: spatial interpolation of highly
variable
data}\label{the-29th-of-june-data-spatial-interpolation-of-highly-variable-data}}

We start estimating the wrapped normal posterior distribution for the
June 29 data. Notice that  distances are computed using the Euclidean distance and hence it is suggested to convert coordinates into UTM system. Computations requires the definition of starting values and prior distributions for the parameters. The latters are:
\begin{itemize}
\item $\alpha \sim WN(\pi, 10)$ (weakly informative).
\item $\rho\sim U(0.0068, 0.0736)$ the interval is chosen by computing practical ranges using the minimum and maximum distances in the data and use them as interval's extremes. 
\item $\sigma^2 \sim IG(3,0.5)$ (informative prior).
\end{itemize}
The covariance is exponential.
Notice that we performed some sensitivity analysis to prior definition. The wrapped model is not very sensitive, a weakly informative prior on $\sigma^2$ will simply require a larger iterations number to achieve  convergence.
The MCMC computation in these examples  runs with  2 chains in parallel allowing the
 use of several convergence checks already available in R (package coda), In
particular we compute the Potential scale reduction factors ($\hat{R}$) of the model
parameters \citep{gelman1992, brooks1998}.

An important practical point is  when to start and end the adaptive procedure of the Metropolis
steps, we can use the command \texttt{start} and \texttt{end} in the
list \texttt{adapt\_param}; \texttt{end} should be
smaller than \texttt{burnin} to ensure that saved samples are drawn from
the desired posterior distribution. The parameter \(\xi_{ad}\) is set
using the command \texttt{exp}, the desired acceptance-ratio \(d_{ad}\)
is \texttt{accept\_ratio} while we assume
\(\boldsymbol{\Sigma}_{ad}^{0}\) to be a diagonal matrix with diagonal
elements given by the \texttt{sd\_prop}. We set 60000 iterations, 30000 of burnin and a thinning of 10. The adaptation starts at the 100th iterations and it ends at iteration 10000.

Graphical checks are obtained using a specific  function \texttt{ConvCheck} and must be read recalling that \texttt{alpha} is a
circular variable
\begin{figure}[t]
\centering
\includegraphics[width=0.70000\textwidth]{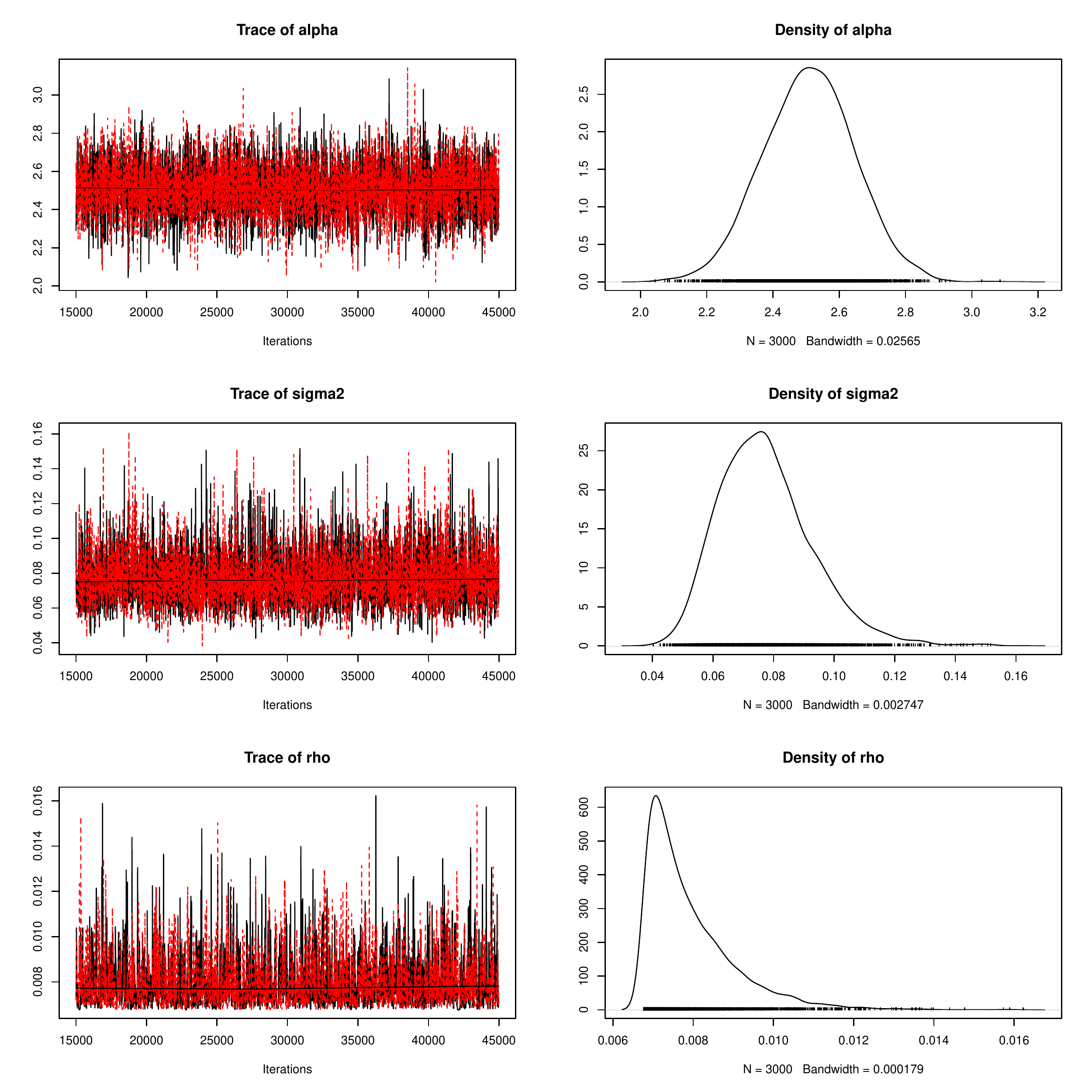}
\caption{Traces and densities from the mcmc run for the wrapped Gaussian spatial model -  June 29, 2012 data}\label{fig:tracce1}
\end{figure}
Convergence is  confirmed by $\hat{R}$s values, all equals to 1. We can proceed to interpolate the
values on the 20 \(\times\) 20 cells grid reported in Figure
\ref{fig:thegrid}. 
Once the predictions are obtained, we compute the APE and the circular
continuous ranked probability score (CRPS). These measures are evaluated
between data points and their nearest grid points. Both functions return
the average index value, and a vector of scores computed on each site. Results are reported in Table \ref{tab:apecrps}.

Now we fit the projected normal model to the same data. Some care must
be taken in setting prior distributions for the projected normal model. This model is more sensitive to the choice of priors  for the
decay parameter. We propose a uniform distribution on a fixed interval, as already mentioned, 
when the decay parameter suggested by the data is close to the extreme of the chosen
interval we may obtain non-positive definite covariance matrices and the
computation will stop. 
\begin{itemize}
\item $\rho\sim U(0.0068, 0.0746)$
\item $\boldsymbol{\alpha}\sim N_2\left(\left( \begin{array}{c} \frac{\pi}{3}\\ \frac{2\pi}{3}\end{array}\right), \left(\begin{array}{cc} 20 & 0\\ 0 & 20 \end{array}\right)\right)$
\item $\tau \sim U(-1,1)$
\item $\sigma^2\sim IG(3,2)$ 
\end{itemize}
 Again the spatial covariance is exponential. With the projected model we run our chains for a longer time (100000 iterations half of them are kept) as the convergence is slower.  The adaptation algorithm differ slightly, as the  standard deviation of the log-normal proposals for \(\mathbf{r}\) must be set.

From the convergence check (including the traceplots that are not
reported) we see that the chains reached convergence. In case an update
is required it can be easily done by setting as starting values the last
values of all chains (see the Supplementary material). We can then move to the prediction step.

\begin{figure}[t]
\centering
\includegraphics[scale = 0.35,trim= 20 20 20 0]{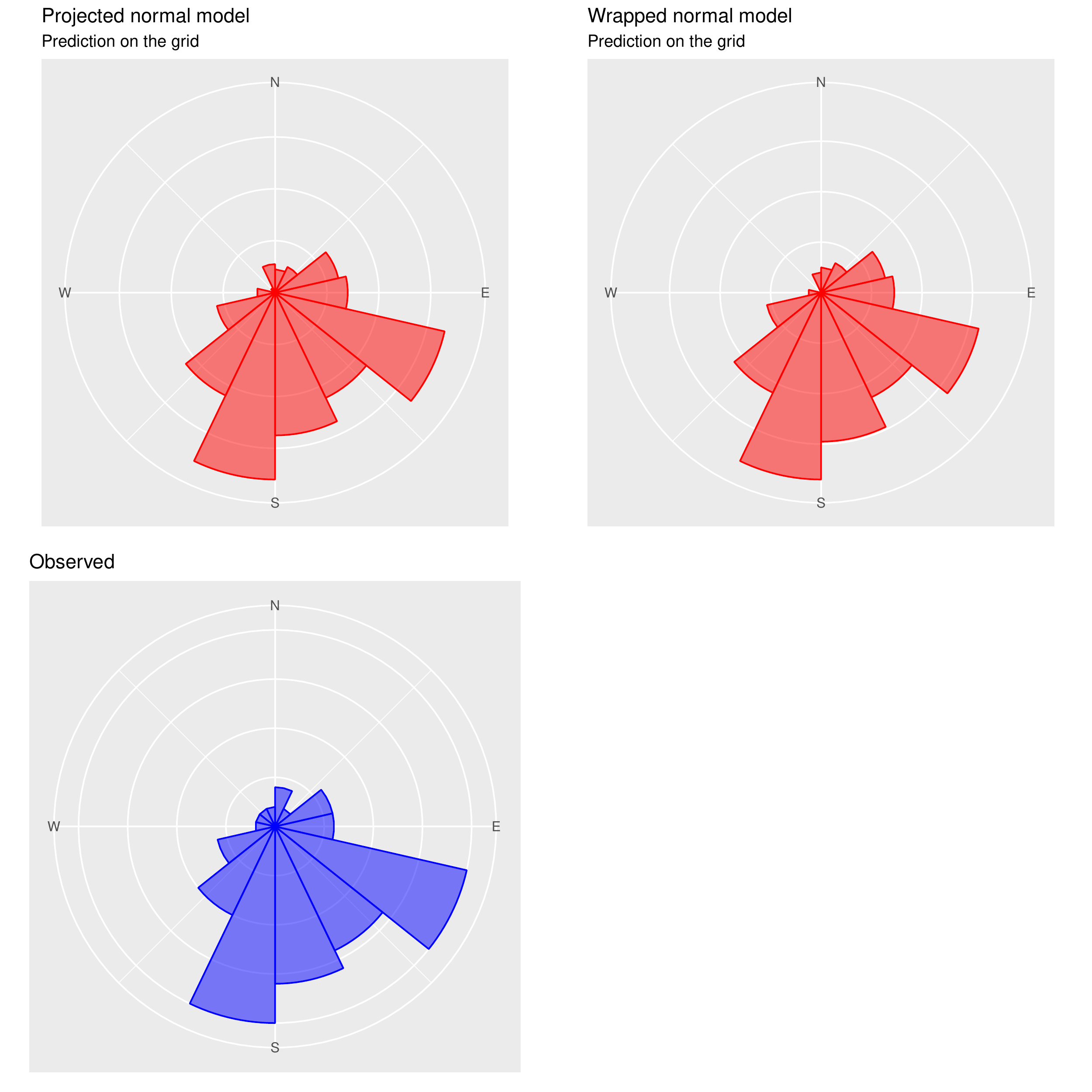}
\caption{Interpolated values (red) and observed values  (blue)}\label{fig:dati1pred}
\end{figure}

\begin{figure}[t]
\centering
\includegraphics[scale=0.3,trim= 20 160 20 100]{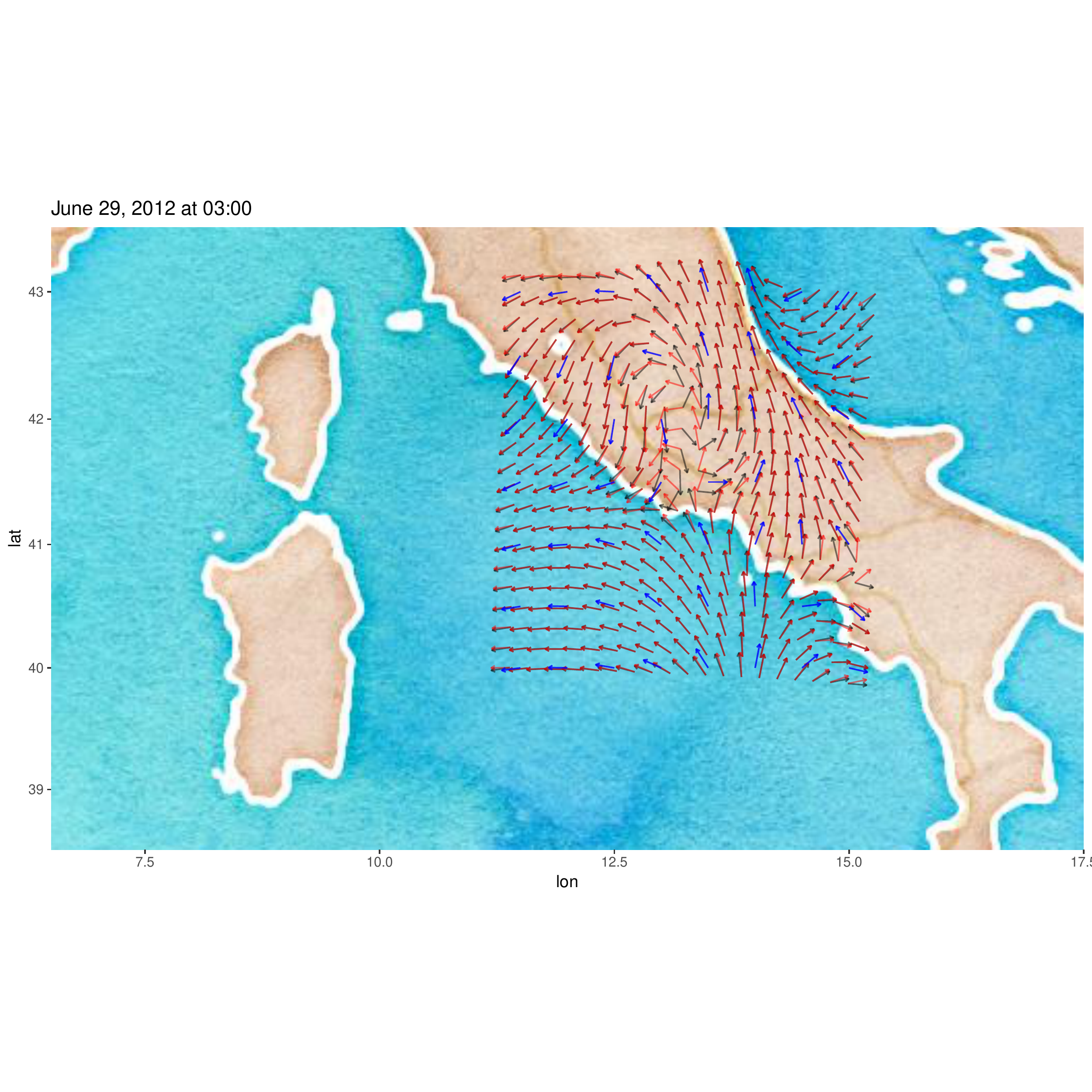}
\caption{June 29,2012. Interpolated values under the wrapped normal model (red), projected normal model (black) together with the observed values  (blue)}\label{fig:dati2pred}
\end{figure}

With the data from June 29, 2012 the projected normal flexibility allows
a better fit of the model (see Table \ref{tab:apecrps} as it is also shown in Figure
\ref{fig:dati1pred}). In Figure \ref{fig:dati2pred} the field
representation of interpolated wind directions are shown together with
the observed values under both models, wrapped and projected,  and
it can be seen as the differences between the two models, however
relevant in terms of APE and CRPS, are hardly visible.

\hypertarget{data-from-the-29th-of-october-2018-spatial-interpolation-of-concentrated-values}{%
\subsection{Data from the 29th of October 2018: spatial interpolation of
concentrated
values}\label{data-from-the-29th-of-october-2018-spatial-interpolation-of-concentrated-values}}

On October 29 and 30 2018, at least 11 people were killed and many
injured after severe storms produced heavy rain and strong winds across
Italy, causing widespread floods and damage. Some of the worst affected
areas were on the west coast, particularly in Liguria. Many people were
evacuated due to increased river levels across the country. Winds up to
180 km/h (112 mph) were reported on October 29, 2018, and blamed for the
deaths of at least 6 of the 11 killed people. Trees were downed from the
country's North down to Rome and further South to Naples. Winds were
mostly from the South-East quadrant (Scirocco).

Storms usually implies
very little variability in the wind direction, then a priori no model
would be preferable. The flexible projected normal would most likely adapt well, but
it will take some time and patience to get estimated values. The WN
would most likely perform well as the distribution of the angle is
unimodal and concentrated (see Figure \ref{fig:bothdataset}).

\begin{figure}[t]
\centering
\includegraphics[scale=0.3,trim= 20 160 20 20]{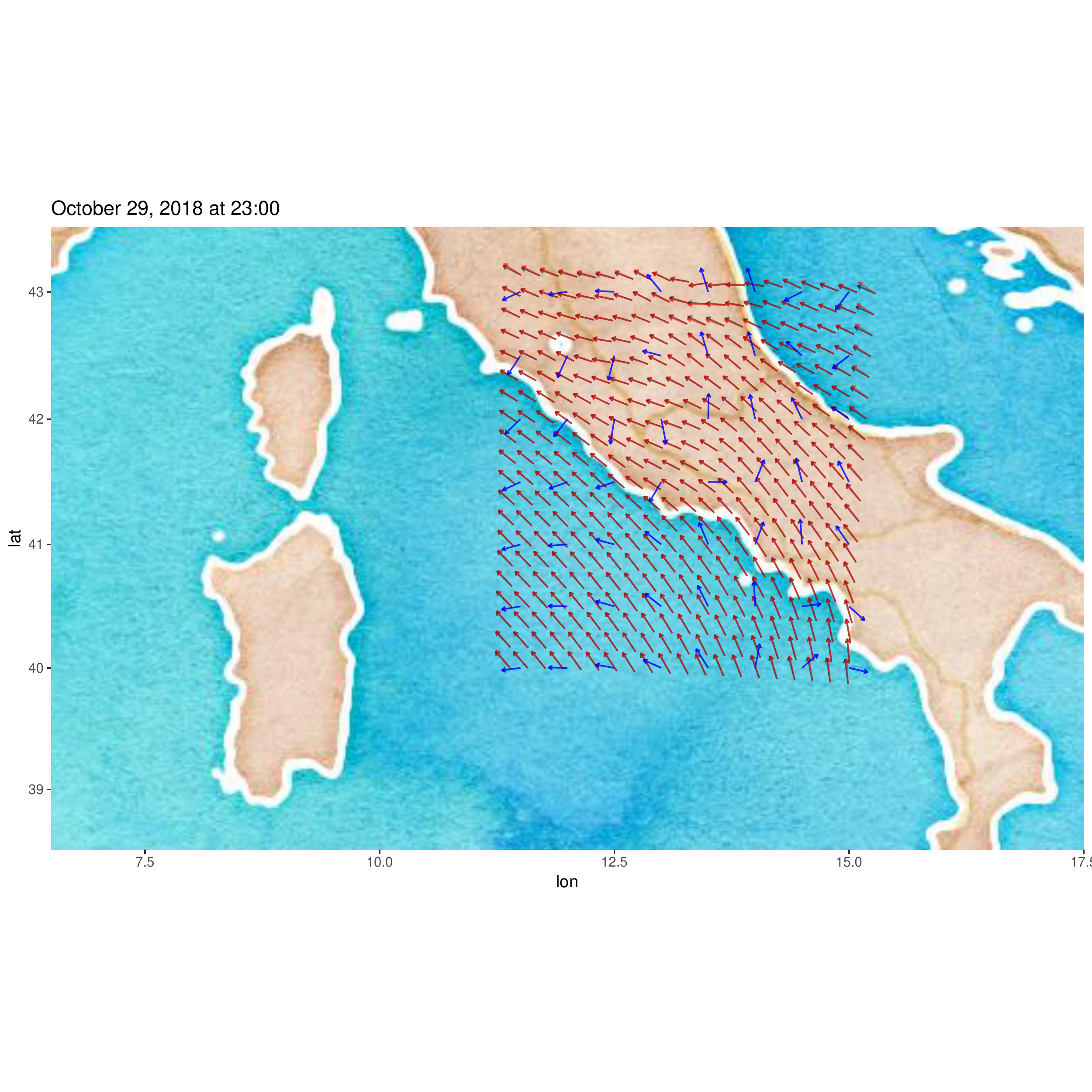}
\caption{October 29, 2018. Interpolated values under the wrapped normal model (red), projected normal model (black) together with the observed values  (blue)}\label{fig:interpottobre}
\end{figure}

We briefly summarize the setting and results. Prior distributions and MCMC settings were:

\begin{itemize}
\item \textbf{Wrapped Normal}
\begin{itemize}
\item $\boldsymbol{\alpha}  \sim N(\frac{2\pi}{3},10)$
 \item $\rho    \sim U(0.0068, 0.0736)$
 \item $\sigma^2 \sim IG(2,0.3)$ we can set an improper prior 
 \item 60000 iterations with 30000 burn-in and thin of 10
 \end{itemize}
 \item \textbf{Projected Normal}
 \begin{itemize}
 \item $\rho\sim U(0.0068, 0.0746)$
  \item $\tau\sim U(-0.5,1)$ we are more informative to speed convergence
\item $\sigma^2\sim IG(6,5)$
\item $\boldsymbol{\alpha}\sim N_2\left(\left( \begin{array}{c} \frac{\pi}{6}\\ \frac{2\pi}{3}\end{array}\right), \left(\begin{array}{cc} 20 & 0\\ 0 & 20 \end{array}\right)\right)$
 \item 100000 iterations with 50000 burn-in and thin of 10
 \end{itemize}
 \item exponential correlation function
 \item 2 chains with parallel computation

\end{itemize}
 The adaptive part is set as in the MCMC described for the June 29 data.
Convergence was achieved and the interpolation computed on the grid,
shown in Figure \ref{fig:thegrid}. The APE and CRPS returned the same
values up to the fourth decimal Figure: 0.0043 APE and 0.0008 CRPS. The
best model for this dataset can be the wrapped or the projected equivalently if the
best fit is the main focus. Wrapped Normal model was computationally
less demanding and allowed the choice of less informative priors,
parameters can be easily interpreted. Hence it can be a preferred
choice. In Figure \ref{fig:interpottobre} we compare graphically the two
prediction it is clear that they are indistiguishable.

\begin{table}[t]
\begin{center}
\begin{tabular}{lcccc}
\hline\hline
Date & Ape PN & Ape WN &CRPS PN& CRPS WN \\\hline
29 June 2012& 0.0499 & 0.0736&  0.0076 &  0.012\\
29 October 2018& 0.0043& 0.0043& 0.0008& 0.0008\\\hline\hline
\end{tabular}
\caption{Spatial data examples: model choice based on APE and CRPS}\label{tab:apecrps}
\end{center}
\end{table}

\hypertarget{spatio-temporal-circular-wrapped-normal-model}{%
\subsection{Spatio-temporal circular wrapped normal
model}\label{spatio-temporal-circular-wrapped-normal-model}}

To illustrate the space-time version of the models implemented in
CircSpaceTime we use  data from the 29th of October 2018
 from 15:00 to 21  every three hours and we predict the time
point at 23:00 that we used in the spatial example. Again we compare the
two modeling approaches. the model complexity increases as well as the computational time.

\begin{figure}[t]
\centering
\includegraphics[scale = 0.35,trim= 20 20 20 20]{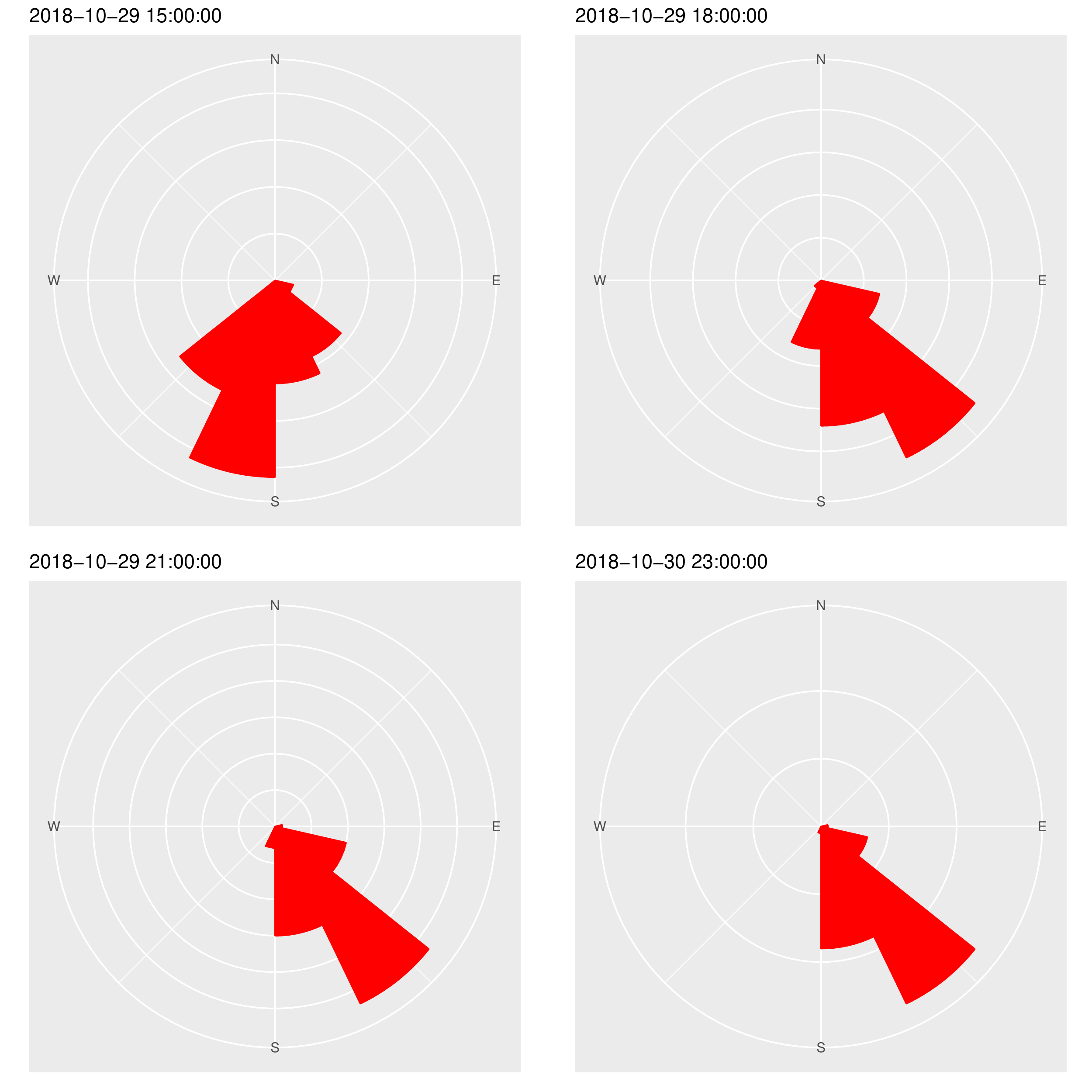}
\caption{Time points from October 29, 2018. The first three are used to predict the 23:00 hour}\label{fig:dati29ott}
\end{figure}

As in the spatial case we set values for the spatial and temporal range
priors 
\begin{itemize}
\item Wrapped Normal spatio-temporal model's settings:
\begin{itemize}
\item $\rho_{sp}\sim U(0.001,0.100)$, $\rho_t\sim U(1, 3)$  the complexity of the model suggests less informative priors for single parameters and hence for  the decay parameters
\item $\eta\sim Beta(1,1)$ 
\item $\sigma^2\sim IG(5,5)$
\item $\alpha \sim WN(\frac{2\pi}{3},20)$
\item iterations 150000, of which 50000 burn-in and we keep one sample out of 10.
\end{itemize}
\item Projected Normal spatio-temporal model's settinngs:
\begin{itemize}
\item $\rho_{sp}\sim U(0.001, 0.574)$, $\rho_t\sim U(0.01, 3.5)$
\item $\eta\sim Beta(1,1)$ 
\item $\sigma^2\sim IG(5,5)$
\item $\tau\sim U(-1,1)$
\item $\boldsymbol{\alpha}\sim N_2\left(\left( \begin{array}{c} \frac{2\pi}{3}\\ \pi\end{array}\right), \left(\begin{array}{cc} 10 & 0\\ 0 & 10 \end{array}\right)\right)$
\item a total of 300000 (two updates) with 150000 burnin and thinning of 10
\end{itemize}
\end{itemize}

The wrapped normal model  convergence check returns a Multivariate Potential Scale reduction Factor ($\hat{R}_M$) of 1.01 after one run and all
parameters chains seem to have reached convergence (not shown).We move to the prediction of the 23:00 hour. The Average prediction error on the predicted locations is 0.16 radians,
ranging from 0.1435 to 0.2258, while the average CRPS is 0.0243, with
point values ranging from 0.0103 to 0.0917.

While the projected normal  model convergence is achieved with multivariate $\hat{R}_M$ equal to 1.05. We move
again to the prediction of the time point at 23:00 and compare the two
models through APE and CRPS. The APE ranges between 0.028 and 0.133 with an average value of 0.068, CRPS ranges between 0.0005 and 0.034, with an average value of  0.008.

\begin{figure}[t]
\centering
\includegraphics[scale=0.35,trim= 20 20 20 20]{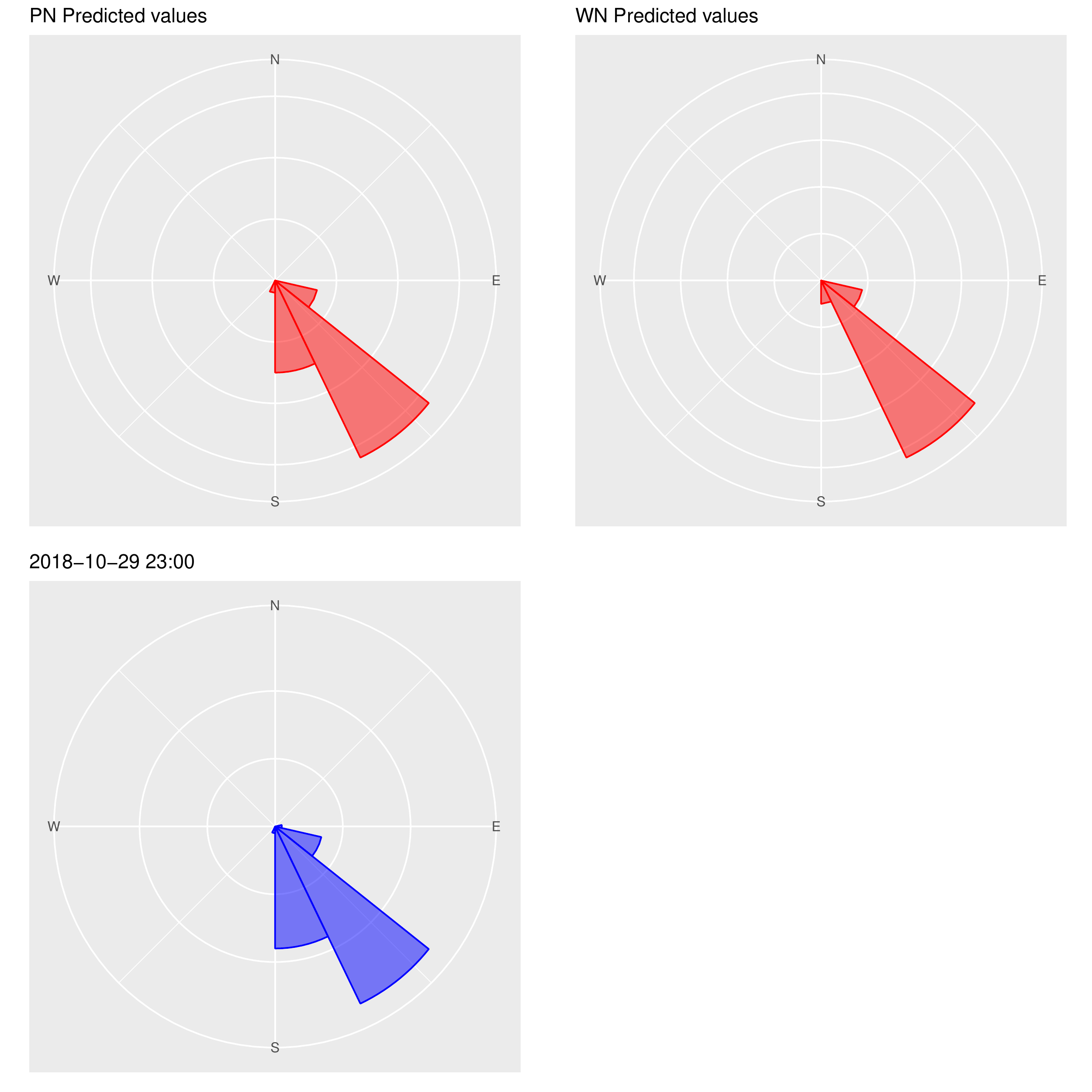}
\caption{Predicted and observed values, 29th of October 2018 at 23:00}\label{fig:comparast}
\end{figure}

It is very clear that the projected normal captures way better the data
features when prediction forward in time is required. In Figure
\ref{fig:comparast} we can see that the wrapped normal model does not capture the second
peak in the wind direction data.

\begin{figure}[t]
\centering
\includegraphics[scale=0.3,trim= 20 160 20 120]{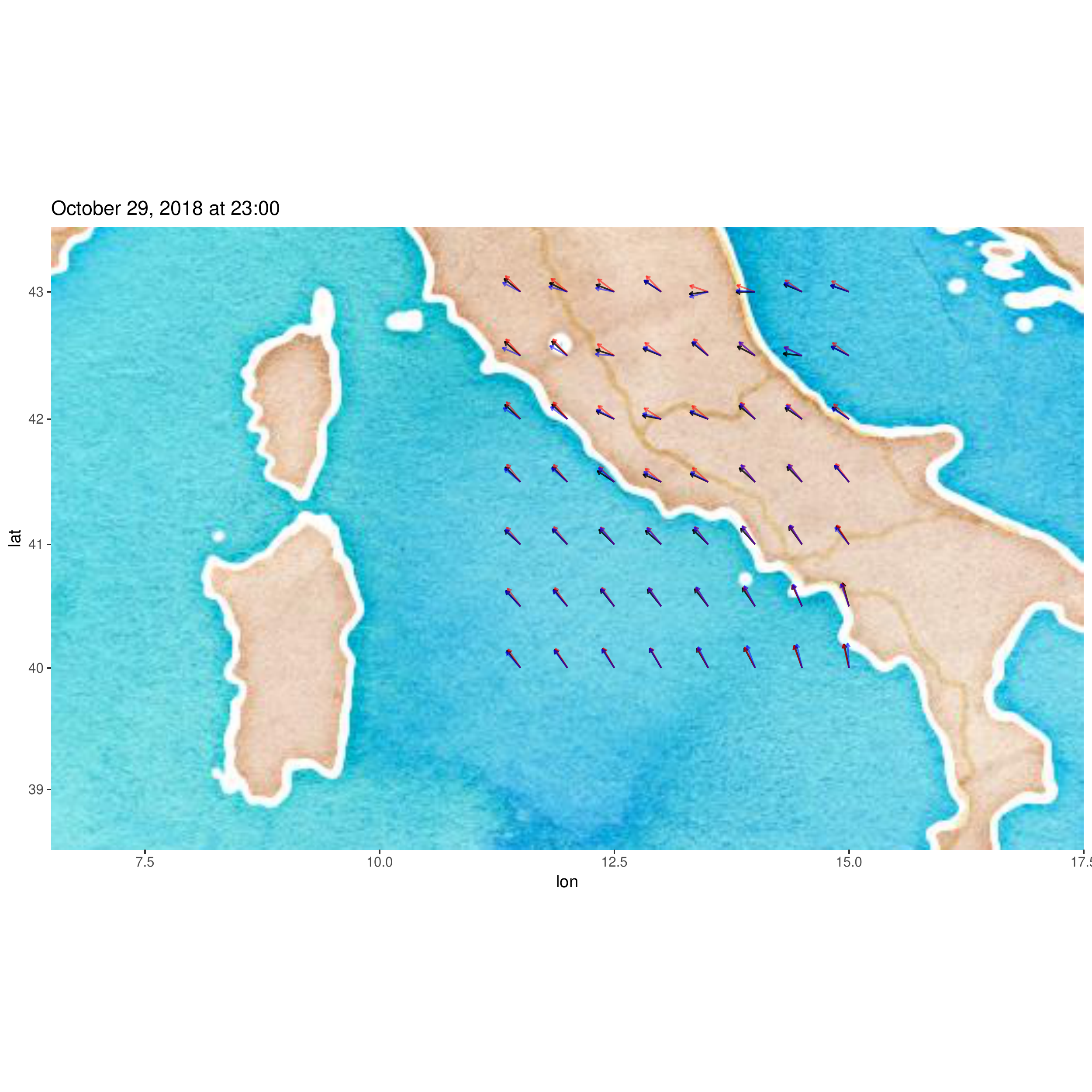}
\caption{Predicted and observed values, 29th of October 2018 at 23:00. Wrapped normal predictions (in red), projected normal predictions (black) and observed values (blue)}\label{fig:comparast2}
\end{figure}

In Figure \ref{fig:comparast2} we can see point by point the behaviour
of the models with respect to the observed values. In the northern
section of the area, where a slightly larger variability is found, the
wrapped normal model produces values further a part from the observed
ones as the projected normal model.

\hypertarget{concluding-remarks}{%
\section{Concluding Remarks}\label{concluding-remarks}}

In this paper we presented the CircSpaceTime package, that collects
functions implementing spatial and spatio-temporal Bayesian models
allowing interpolation of dependent circular data. The implemented
models have been introduced in a series of papers that are briefly
summarized here using a new formalization
\citep[see][for details]{Jona2013, Wang2013, wang2014, mastrantonio2015b}.
The current version of the package does not include models with spatial nugget
and allows for constant mean only. Future versions will include nugget
and regression-type mean. We also plan to include more models for
dependent circular data spreading from the works \citet{Bulla2012},
\citet{lagona2012}, \citet{bulla2015}, \citet{lagona2015} and
\citet{Maruotti2016}, that develop likelihood-based estimation for
complex modeling of circular and cylindrical data.

\section*{Acknowledgments}

This work has been partially developed under the PRIN2015
supported-project \textit{Environmental processes and human
activities: capturing their interactions via statistical methods}
(EPHAStat) funded by MIUR (Italian Ministry of Education, University and
Scientific Research) (20154X8K23-SH3). Gianluca Mastrantonio research
has been partially supported by MIUR grant Dipartimenti di Eccellenza
2018 - 2022 (E11G18000350001).

\bibliographystyle{tfcad}
\bibliography{CircRef.bib}

\end{document}